\documentclass[useAMS,usenatbib]{mnras}
\usepackage{graphicx}
\usepackage[section]{placeins}

\usepackage{yfonts}
\usepackage{times} 
\usepackage{amsmath}
\usepackage{amssymb}
\usepackage{tikz}
\usepackage{xcolor}

\def\icarus{Icarus}

\def\aj{Astron.\ J. }
\def\apj{Astrophys.\ J. }
\def\apjl{Astrophys.\ J. }
\def\apjs{Astrophys.\ J.\ (Supp.) }
\def\aap{Astron.\ Astrophys. }

\def\bssA_{Bull.\ Seismol.\ Soc.\ Am. }

\def\gcA_{Geochim.\ Cosmochim.\ Acta }

\def\mnras{Mon.\ Not.\ R.\ Astron.\ Soc. }

\def\planss{Plan.\ Space Sci. }
\def\prA_{Phys.\ Rev.\ A }

\begin{document}

\title[Origin of Asteroid (514107) 2015 BZ509]{An interstellar origin for Jupiter's retrograde co-orbital asteroid}
\author[F. Namouni and M. H. M. Morais]{F. Namouni$^{1}$\thanks{E-mail:
namouni@obs-nice.fr (FN) ; helena.morais@rc.unesp.br (MHMM)} and  M. H. M. Morais
$^{2}$\footnotemark[1]\\
$^{1}$Universit\'e C\^ote d'Azur, CNRS, Observatoire de la C\^ote d'Azur, CS 34229, 06304 Nice, France\\
$^{2}$Instituto de Geoci\^encias e Ci\^encias Exatas, Universidade Estadual Paulista (UNESP), Av. 24-A, 1515 13506-900 Rio Claro, SP, Brazil}

\date{Accepted 2018 March 26. Received 2018 March 23; in original form 2018 February 26.}

\maketitle

\begin{abstract}
Asteroid (514107) 2015 BZ509 was discovered recently in Jupiter's co-orbital region with a retrograde motion around the Sun. The known chaotic dynamics of the outer Solar System have so far precluded the identification of its origin. Here, we perform a {high-resolution} statistical search for stable orbits and show that asteroid (514107) 2015 BZ509 has been in its current orbital state since the formation of the Solar System. This result indicates that  (514107) 2015 BZ509 was captured from the interstellar medium 4.5 billion years in the past as planet formation models cannot produce such a primordial large-inclination orbit with the planets on nearly-coplanar orbits interacting with a coplanar debris disk that must produce the low-inclination small-body reservoirs of the Solar System such as the asteroid and Kuiper belts. This result also implies that more extrasolar  asteroids are currently present in the Solar System on nearly-polar orbits.  
\end{abstract}


\section{Introduction}\label{section1}
Centaurs, the asteroids that roam the space between the giant planets of the Solar System, have a chaotic dynamical evolution governed for some by the close encounters with the giant planets, and for others by their successive hopping in and out of the outer planetsÕ web of mean motion resonances. The mean lifetime of the first group ranges from 1 to 10 million years whereas that of the second group extends to 100 million years as their resonant status provides them with some dynamical protection \citep{TiscarenoMalhotra03,BaileyMalhotra09,VolkMalhotra13}. Capture in resonance may occur for prograde or retrograde motion with a greater likelihood for the latter making the retrograde resonant Centaurs possibly the oldest asteroid residents of the outer Solar System \citep{NamouniMorais15,NamouniMorais17}. A number of Centaurs are currently known to be in retrograde resonance between the outer planets' orbits \citep{MoraisNamouni13b} but it is the discovery of asteroid (514107) 2015 BZ509, inside Jupiter's co-orbital region, with an orbit of moderate eccentricity of 0.38 and a retrograde inclination of 163$^\circ$ \citep{Wiegert17} that has produced so far the most puzzling example of retrograde resonance in the Solar System. 

Jupiter's co-orbital region is known to host the Trojan asteroids that were captured mostly permanently by the planet during the late stage of Solar System formation \citep{Tsiganis05,Robutel06,Nesvorny13,Jewitt18}. Asteroid 2015 BZ509 shares the co-orbital region with the Trojans but moves in the opposite orbital direction. It sits on Jupiter's peak of capture probability \citep{NamouniMorais17c} and would have an indefinitely stable orbit if the Solar System contained only Jupiter \citep{MoraisNamouni13a,MoraisNamouni16}. It is thought to be trapped temporarily as standard planet formation models do not produce Centaurs in situ with long term stable retrograde orbits. In this framework, 2015 BZ509 could originate from distant reservoirs such as the scattered disk or the Oort cloud like typical Centaurs \citep{Brasser12}. To reach Jupiter's orbit, 2015 BZ509 would have had to cross the giant planets' space whose chaotic dynamics and its 100-million-year lifetime timescale suggest a recent capture at Jupiter's orbit and preclude the identification of the asteroid's origin during the final stage of planet formation some 4 billion years ago. 

However, when the motion of 100 asteroid clones with orbits that differ slightly from that of 2015 BZ509 by amounts compatible with the orbital elements' error bars, was simulated over one million years, it was found to have a stable evolution \citep{Wiegert17}. This timescale is about two orders of magnitude longer than those of temporarily captured retrograde resonant asteroids  \citep{MoraisNamouni13b}  hinting to a different origin from that of most Centaurs.

In this Letter, we report on a high-resolution statistical search for stable orbits for 2015 BZ509 that allows us to trace the asteroid's origin back to the epoch of Solar System formation. In Section 2, we describe our numerical approach of simulating the evolution of one million clones of 2015 BZ509. In Section 3, we present  our results that show 2015 BZ509 has been a co-orbital of Jupiter since the end of planet formation and has a strongly stable orbit that can live theoretically {at least 43 billion years}. In Section 4, we explain that this finding implies that 2015 BZ509 was captured from the interstellar medium and that there should be more extrasolar asteroids bound to the Solar System on nearly-polar orbits.

\begin{table}
\centering
\caption{Nominal orbital elements of asteroid (514107) 2015 BZ509 and their standard deviations.}
\label{table:1}
\begin{tabular}{|l|r}
\hline \hline
Julian date & $2457800.5$\\
Semi-major axis (au)& $5.140344420\pm 5.7871 \times 10^{-5}$\\
Eccentricity & $0.3806619168 \pm 6.2258 \times 10^{-6}$\\
Inclination ($^\circ$)& $163.004558786\pm 2.0009 \times 10^{-5}$\\
Longitude of ascending node ($^\circ$)&$307.376837106 \pm3.5503 \times 10^{-5}$\\
Argument of perihelion ($^\circ$)&$257.44919796 \pm 2.7616 \times10^{-4}$\\
Mean anomaly ($^\circ$)&$32.46169240 \pm 4.4177 \times 10^{-4 }$\\
\hline \hline
\end{tabular}
\end{table}

\section{One million clone simulation}\label{section2}
The motion of the asteroid and the planets constitutes an $N$-body dynamical system whose phase space structure is complex and chaotic but does not preclude the presence of long term stability islands even at or between the outer giant planets. Probing phase space to identify such islands requires a significant number of initial conditions especially since 2015 BZ509's precise orbit is unknown and only a representation thereof in parameter space by a nominal orbit and a covariance matrix exists \citep{Knezevic12}. We therefore simulated the evolution of one million asteroid clones that interact with the giant planets and the Galactic tide back to 43 billion years in the past to search for the most stable orbits. {The existence of stable orbits over the age of the Solar System is by no means guaranteed especially as we probe one of the most chaotic regions in the Solar System. However if such orbits exist then they are the ones that correspond to the actual motion of the asteroid and not the short-lived unstable orbits.  Choosing the former orbits over the latter is motivated by the copernican principle that 2015 BZ509 is not being observed at a preferred epoch in Solar System history. }

\begin{figure}
\begin{center}
{ 
\hspace*{-21mm}\includegraphics[width=135mm]{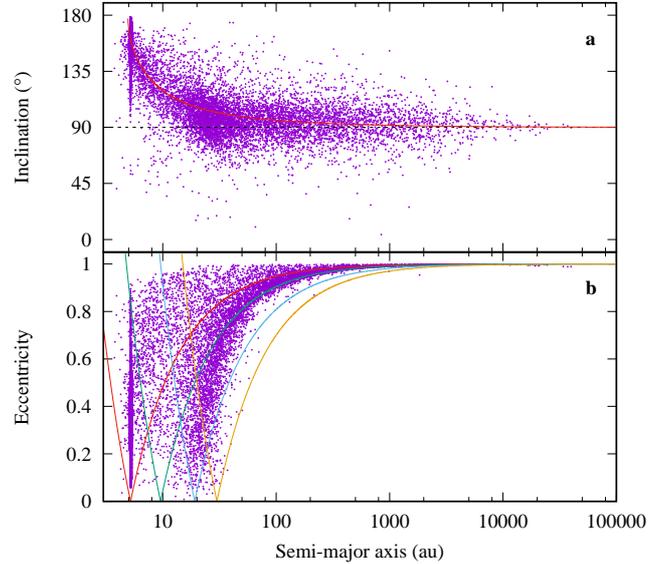}\\[-10mm]
}
\caption{Distribution of 2015 BZ509's clones at 40 million years in the past  in the semi-major axis-inclination plane (a) and the semi-major axis-eccentricity plane (b). The curve in (a) is given by the asteroid's Tisserand relation with Jupiter. The four V-shaped curves in (b) denote the intersection at aphelion and perihelion of the clone's orbit with those of Jupiter, Saturn, Uranus and Neptune respectively from left to right.}\label{f1}
\end{center}
\end{figure}

The nominal orbit of 2015 BZ509 (Table 1) and its equinoctial covariance matrix were obtained from the AstDys database\footnote{http://hamilton.dm.unipi.it/astdys \citep{Knezevic12}}  for the Julian date 2457800.5. The orbital elements of the planets were obtained from NASA JPL's Horizons ephemeris system\footnote{http://ssd.jpl.nasa.gov \citep{Giorgini96}} for the same epoch. Clone orbits were generated using the Cholesky method for multivariate normal distributions \citep{cholesky}. This method consists in writing the equinoctial covariance matrix as $C = L L^t$ where $L$ is a lower triangular matrix. A sample of one million clones is generated from the equinoctial nominal elements $e_{0i}$  where $1 \leq i \leq 6$ as $e_i =e_{0i} + r_j L_{ij}$; summation over the repeated $j$ index is implied with $j \leq i$, and $r_j$ is a 6-dimensional vector with components generated independently from a normal distribution with mean 0 and variance 1. The one million sample achieves $10^{-7}$ to $10^{-5}$ relative error in reproducing the observational covariance matrix whereas smaller samples achieve larger relative error (e.g. $10^{-2}$ to $10^{-1}$ for 1000 clones) because of the large dimension of the initial conditions' space.

The evolution of an asteroid clone back in time was followed in the system composed of the four giant planets and the Sun whose mass was augmented by those of the inner Solar System's planets. The full three-dimensional Galactic tide \citep{Heisler86} and relative inclination of the ecliptic and Galactic planes were taken into account. The Oort constants ($A=15.3$ km\,s$^{-1}$kpc$^{-1}$, $B=-11.9$ km\,s$^{-1}$kpc$^{-1}$) and star density in the solar neighborhood ($\rho_0 = 0.119 M_\odot$pc$^{-3}$) were taken from the recent Gaia DR1 determinations \citep{Bovy17,Widmark17}. The five-body problem with the Galactic tide is adequate to analyze the stability of 2015 BZ509 as stable clones surviving in the co-orbital region near the nominal orbit have a perihelion at 3.6 au far outside Mars's orbit thus precluding close encounters with the inner Solar System's planets. Numerical integration was carried out using the Bulirsch and Stoer algorithm {with an error tolerance of $10^{-11}.$ More standard symplectic-based alternatives such as the hybrid Mixed-Variable Symplectic integrator (MVS) \citep{Chambers99} were not used because of  the peculiar geometry of the retrograde co-orbital resonance that implies the asteroid encounters the planet twice per period. Tests with the hybrid MVS show that that code switches to the Bulirsch and Stoer algorithm on a large portion of the orbit because of that geometry. To avoid such systematic and frequent algorithm switching, we opted for the Bulirsch and Stoer algorithm as it is also adequate to model large eccentricity orbit evolution \citep{Wiegert99}.} Orbital evolution was monitored for the following events: collision with the Sun, collision with the planets, ejection from the Solar System, and reaching the inner 1 au semi-major axis boundary. No event at the inner boundary was registered.

\section{Long-term stable orbits}\label{section3}
The simulation shows that the clone minimum and median lifetimes are respectively 0.29 million years and 6.48 million years. Unstable clones that exit the co-orbital region undergo close encounters with the planets and are temporarily captured in mean motion resonances much like the general behavior of Centaurs. They also follow a distinct path towards polar inclinations into a dynamical structure that extends to the Oort cloud and that we term `the polar corridor', before being removed from the system. Fig. 1 illustrates this structure with a system snapshot after an evolution back in time of 40 million years when 43628 clones are present. {The dynamical structure starting from the current location of 2015 BZ509 is centered around the curve of the asteroid's Tisserand relation with Jupiter. The clones follow this curve in the first few million years of evolution only to be dispersed around it by their encounters with the other giant planets.} The location breakdown at 40 million years is as follows: 35372 in retrograde co-orbital resonance with Jupiter, 52  in the inner Solar System, 3361  between Jupiter's and Neptune's orbits, 4296 trans-Neptunians with semi-major axes smaller than 1000 au and 547 with semi-major axes larger than 1000 au that extend to the Oort cloud.

\begin{figure}
\begin{center}
{ 
\hspace*{-21mm}\includegraphics[width=135mm]{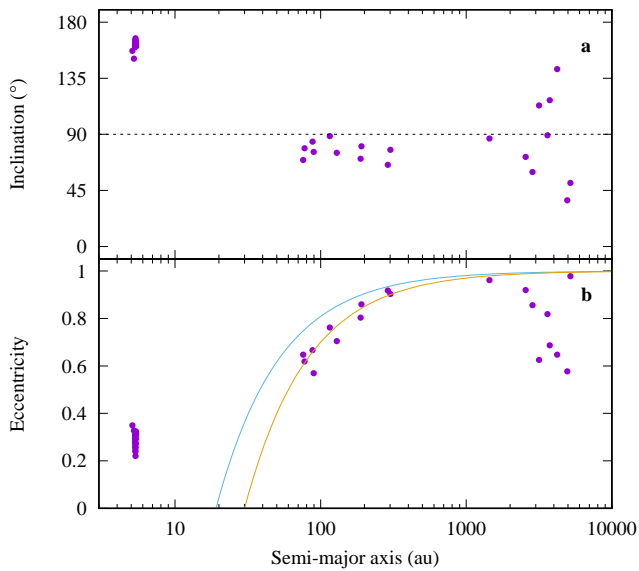}\\[-10mm]
}
\caption{Distribution of 2015 BZ509's clones at 4.5 billion years in the past in the semi-major axis-inclination plane (a) and the semi-major axis-eccentricity plane (b). The intersections at perihelion with the orbits of
Uranus and Neptune are shown by the left and right curves respectively. }\label{f2}
\end{center}
\end{figure}

\begin{figure}
\begin{center}
{ 
\hspace*{-19.5mm}\includegraphics[width=133mm]{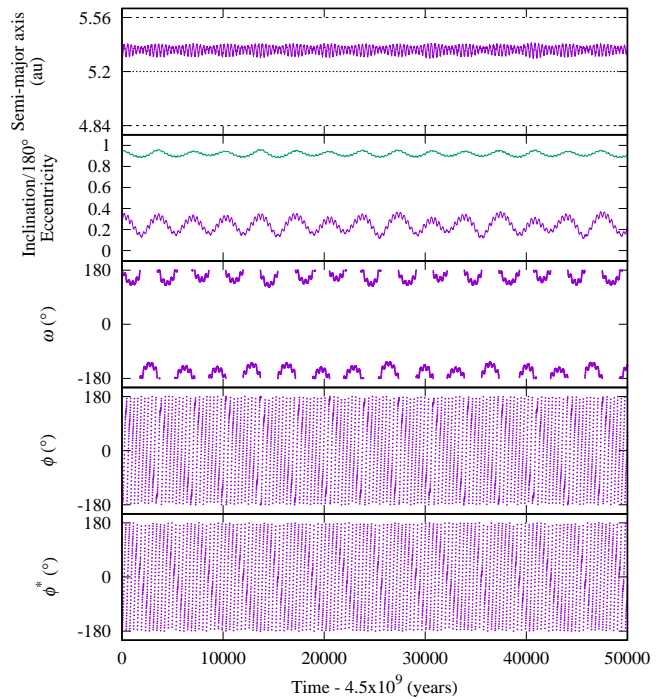}
}
\caption{Evolution of a stable co-orbital clone's orbital elements near 4.5 billion years in the past. The location of the semi-major axis is shown with respect to Jupiter's coorbital region (dashed lines). Inclination is normalized to $180^\circ$ and corresponds to the top curve in the corresponding panel whereas eccentricity is the bottom curve. The possible resonant arguments $\omega$, $\phi$, and $\phi^\star$ are defined in Section 3.}\label{f3}
\end{center}
\end{figure}

At the 100-million-year Centaur maximum instability timescale, 6577 clones remain stable, 75 per cent of which are sheltered by Jupiter's co-orbital resonance. At the end of planet formation, 4.5 billion years in the past, most clones are lost: 553811 increased their eccentricities to unity thereby reaching the Sun's surface, 445678 were ejected from the Solar System and 465 collided with a planet. The surviving clones number 46 of which 27 are in co-orbital resonance with Jupiter whereas the remaining 19 are dispersed between the current locations of the scattered disk and the inner Oort cloud as shown in Fig. 2. The 10 clones in the scattered disk region reside in the polar corridor and have high inclination prograde orbits with perihelia at Uranus's or Neptune's orbits. Most of the 9 clones in the Oort cloud region had their orbits extracted from the polar corridor by the Galactic tide and do not suffer close encounters with the planets at the corresponding epoch. 
 
\begin{figure*}
\begin{center}
{ \vspace{-46mm}\hspace*{-4mm}\includegraphics[width=193mm]{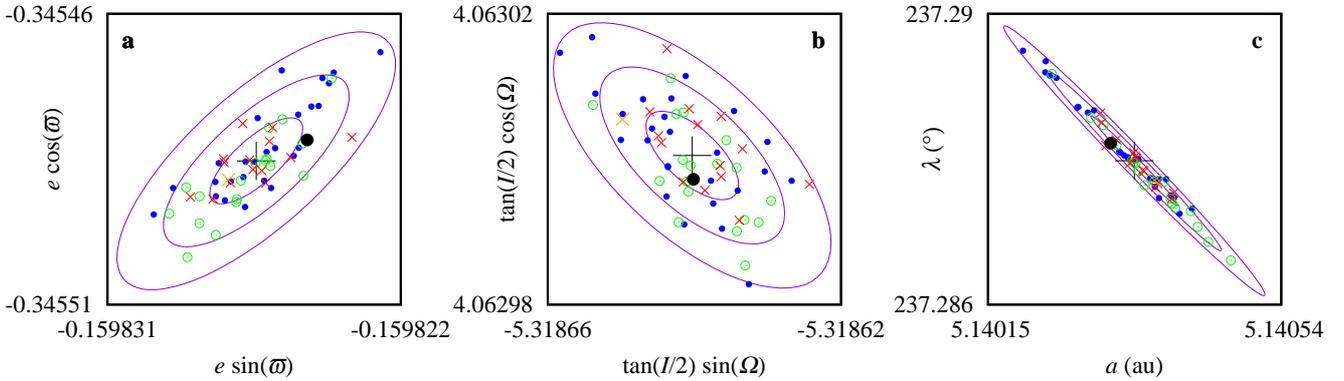}\\[-25mm]
}
\caption{Distribution of the 61 initial conditions of 2015 BZ509Õs clones  that were present from 3.5 to 4.5 billion years in the past. Panel (a) shows the initial conditions in the ($e\sin\varpi$,$e\cos\varpi$) plane where $e$ is the eccentricity and $\varpi$ is the longitude of perihelion. Panel (b) shows the initial conditions in the ($\tan (I/2)\sin\Omega$,$\tan(I/2)\cos\varpi$) plane where $I$ is the inclination and $\Omega$ is the longitude of the ascending node. {Panel (c)} shows the initial conditions in the ($a$,$\lambda$) plane where $a$ is the semi-major axis $\lambda$ is the mean longitude. In each panel the 1-$\sigma$, 2-$\sigma$ and 3-$\sigma$ confidence levels of the covariance matrix are shown around the nominal orbit (large central cross). A clone's state is shown at 4.5 billion years in the past: co-orbital clones  with blue filled circles,  non-co-orbital clones with empty green circles, ejected clones with small red crosses and the only Sun-colliding clone with a medium orange cross. {The location of the longest-lived clone (42.91 billion years) is denoted by a large black filled circle.}}\label{f4}
\end{center}
\end{figure*}

The clustering of 60 percent of long term stable clones at 4.5 billion years in Jupiter's co-orbital region shows that 2015 BZ509 has been in its current state since planet formation. For the clones in the current regions of the Oort cloud and the scattered disk, all semi-major axes are widely spaced and have each a 2 percent chance of being the original semi-major axis. 

The 27 long-term stable co-orbital clones share a number of orbital properties regarding their final states at 4.5 billion years illustrated by the example shown in Fig.3. First, all but one clone have increased their semi-major axis above Jupiter's. The average orbital elements and their standard deviations are: semi-major axis $5.3319$\,au\,$ \pm 0.0531$\,au, eccentricity $0.2888\pm 0.0294 $ and inclination $161.9939^\circ \pm 3.0144^\circ$. Second, none of the clones librate with the 1:1 resonance arguments $\phi=\lambda-\lambda_{\rm Jupiter}$ or $\phi^\star=\lambda-\lambda_{\rm Jupiter}-2\omega$ where  $\lambda$, $\lambda_{\rm Jupiter}$ and $\omega$ are respectively the mean longitudes of the clone and Jupiter and the clone's argument of perihelion \citep{MoraisNamouni13a}. Third, they are all solidly trapped in the Kozai-Lidov secular resonance with $\omega=0^\circ$ or $180^\circ$\footnote{The values $\omega=0^\circ$ and $180^\circ$  represent the same equilibrium as the Kozai-Lidov resonance is 180$^\circ$-periodic. {Note  that the eccentricity and inclination oscillate with the same phase because motion is retrograde. In the Kozai-Lidov cycle, eccentricity is maximal when motion is nearly co-planar.  For retrograde motion this occurs when inclination is near $180^\circ$ \citep{NamouniMorais17b}.}}  \citep{Kozai62,Lidov62,MoraisNamouni16}. 

The phase space structure near 2015 BZ509's orbit may be visualized through the distribution of the 61 clones' initial conditions that were present in the first billion years after planet formation (i.e. in the interval [$-4.5$:$-3.5$] billion years).\footnote{We do not show similar data for a more recent interval to avoid an overcrowded plot.}  At 3.5 billion years in the past, there were 36 clones in the co-orbital region, 15 in the scattered disk region and 10 in the Oort cloud region.   The distribution is shown in Fig. 4 with respect to the nominal orbit and the confidence levels of the covariance matrix according to the clone's state at 4.5 billion years in the past.  {The clones do not cluster in terms of co-orbital orbits, non-co-orbital ones, ejected or Sun-colliding orbits indicating, in mathematical terms, that the set of long-term stable orbits and the set of unstable orbits are dense in parameter space around the nominal orbit. In physical terms, this means that a stability island does exist but different nearby orbits within it have different lifetimes.} Dynamical instability and removal from the co-orbital region are likely caused by a slow chaotic diffusion  {from the secular and near-mean motion resonances} similar to that of the Trojan asteroids  \citep{Tsiganis05,Robutel06}. 

Running the simulation further back in time to test orbital stability shows that the clones in the co-orbital region, scattered disk and Oort cloud number (14, 3, 8) at 9 billion years,  (8, 3, 5) at 14 billion years, and (2, 2, 1) at 30 billion years. {At 42.91 billion years, the last co-orbital clone leaves Jupiter's  orbit with a stable motion into the scattered disk region. The slow number decay of stable co-orbitals is  further indication of possible chaotic diffusion. However, on such very long integration timespans, accumulation of numerical error could contribute to the clone's earlier exit from the co-orbital region. That is why the 42.91 billion year estimate is likely a lower bound on the clone's lifetime. A finer analysis of the dynamics in the co-orbital region is required to ascertain the various diffusion times associated with the secular and near-mean motion resonances as well as the role of numerical error in the asteroid's evolution on very long time scales.} The decreasing number of Oort cloud clones is caused by the Galactic tide as it conserves the vertical component of angular momentum forcing slow cyclic oscillations of the clone's eccentricity and inclination that lead to collisions with the planets \citep{Heisler86}.

\section{Interstellar origin}\label{section4}

The presence of (514107) 2015 BZ509 in retrograde co-orbital resonance early in the Solar System's timeline is unexpected from the standpoint of Solar System formation theory as retrograde Centaurs are believed to originate in the scattered disk or the Oort cloud well  after the planets settled down dynamically \citep{Brasser12}. Furthermore, planet formation models cannot produce such a primordial large inclination orbit as that of 2015 BZ509 with the planets on nearly-coplanar orbits interacting with a coplanar debris disk that must produce the low inclination small body reservoirs of the Solar System such as the asteroid and Kuiper belts \citep{Pfalzner15}. This implies that 2015 BZ509 was captured from the interstellar medium.   
{In this respect, even the  low-probability long-lived orbits found in the scattered disk and Oort cloud regions (Fig. 2) should have an interstellar origin because of their location and high inclinations at the end of planet formation}.  
Interstellar capture events can occur during planet formation in a tightly-packed star cluster whose relaxation was more violent than the one that was thought to have formed the Oort cloud in the Sun's birth cluster early in the Solar System's history \citep{Levison10} as the smallest captured semi-major axis was only 1100 au.

The one million clone simulation provides further evidence that there are currently more extrasolar asteroids in the Solar System. In effect,  if more objects were captured along 2015 BZ509 by Jupiter early in the Solar System's history, the less stable orbits must have left the coorbital region by way of chaotic diffusion into the polar corridor. This occurs because the $N$-body problem is time-reversible and unstable clones of 2015 BZ509 that are followed into the future exit the co-orbital region and end up in the polar corridor. The prominent presence of the polar corridor in the simulation over the age of the Solar {System} mainly in the trans-Neptunian region implies that it is currently populated by extrasolar asteroids.  Interestingly, a structure similar to the polar corridor was observed in the known nearly-polar trans-Neptunian objects (TNOs) \citep{Gladmanetal09,Chen16,MoraisNamouni17b}. Integrations of a 1000 clones of TNOs (471325) and 2008 KV42 over 1 billion years  have shown their orbits to evolve toward larger semi-major axes while they cluster around 90$^\circ$ inclination (see Fig. 1 in \cite{Chen16}). 

The presence of extrasolar asteroids  bound to the Solar System early in its history implies the need for a revision of planet formation theory as such interstellar contamination of small body reservoirs will affect not only the dynamics of small bodies but also their physical properties. Observed discrepancies such as that of Trojan colors may originate in a different Trojan origin at different planets \citep{Jewitt18}.

{The stability search method presented here is new and is aimed at beating dynamical unpredictability in one the most chaotic regions of the Solar System using large statistics and intensive computing. It is similar in principle to the orbit determination of newly discovered multi-planet systems that are systematically vetted for stability upon discovery and only the stable orbits are chosen from the available range of observational error bars. In both problems, a stable configuration is preferable to an unstable one as the opposite would imply that the system is being observed at a preferred epoch. Applying systematically our new method to  Centaurs and TNOs will help constrain their origin and improve our understanding of Solar System formation. }

\section*{Acknowledgments}
We thank the anonymous reviewer for helpful comments. The numerical simulation was done at the M\'esocentre SIGAMM hosted at the Observatoire de la C\^ote d'Azur.  M.H.M.M. was supported by grant 2015/17962-5 of S\~ao Paulo Research Foundation (FAPESP)
\bibliographystyle{mnras}

\bibliography{ms}

\begin{thebibliography}{}
\makeatletter
\relax
\def\mn@urlcharsother{\let\do\@makeother \do\$\do\&\do\#\do\^\do\_\do\%\do\~}
\def\mn@doi{\begingroup\mn@urlcharsother \@ifnextchar [ {\mn@doi@}
  {\mn@doi@[]}}
\def\mn@doi@[#1]#2{\def\@tempa{#1}\ifx\@tempa\@empty \href
  {http://dx.doi.org/#2} {doi:#2}\else \href {http://dx.doi.org/#2} {#1}\fi
  \endgroup}
\def\mn@eprint#1#2{\mn@eprint@#1:#2::\@nil}
\def\mn@eprint@arXiv#1{\href {http://arxiv.org/abs/#1} {{\tt arXiv:#1}}}
\def\mn@eprint@dblp#1{\href {http://dblp.uni-trier.de/rec/bibtex/#1.xml}
  {dblp:#1}}
\def\mn@eprint@#1:#2:#3:#4\@nil{\def\@tempa {#1}\def\@tempb {#2}\def\@tempc
  {#3}\ifx \@tempc \@empty \let \@tempc \@tempb \let \@tempb \@tempa \fi \ifx
  \@tempb \@empty \def\@tempb {arXiv}\fi \@ifundefined
  {mn@eprint@\@tempb}{\@tempb:\@tempc}{\expandafter \expandafter \csname
  mn@eprint@\@tempb\endcsname \expandafter{\@tempc}}}

\bibitem[\protect\citeauthoryear{{Bailey} \& {Malhotra}}{{Bailey} \&
  {Malhotra}}{2009}]{BaileyMalhotra09}
{Bailey} B.~L.,  {Malhotra} R.,  2009, \mn@doi [\icarus]
  {10.1016/j.icarus.2009.03.044}, \href
  {http://adsabs.harvard.edu/abs/2009Icar..203..155B} {203, 155}

\bibitem[\protect\citeauthoryear{{Bovy}}{{Bovy}}{2017}]{Bovy17}
{Bovy} J.,  2017, \mn@doi [\mnras] {10.1093/mnrasl/slx027}, \href
  {http://adsabs.harvard.edu/abs/2017MNRAS.468L..63B} {468, L63}

\bibitem[\protect\citeauthoryear{{Brasser}, {Schwamb}, {Lykawka}  \&
  {Gomes}}{{Brasser} et~al.}{2012}]{Brasser12}
{Brasser} R.,  {Schwamb} M.~E.,  {Lykawka} P.~S.,   {Gomes} R.~S.,  2012,
  \mn@doi [\mnras] {10.1111/j.1365-2966.2011.20264.x}, \href
  {http://adsabs.harvard.edu/abs/2012MNRAS.420.3396B} {420, 3396}

\bibitem[\protect\citeauthoryear{{Chambers}}{{Chambers}}{1999}]{Chambers99}
{Chambers} J.~E.,  1999, \mn@doi [\mnras] {10.1046/j.1365-8711.1999.02379.x},
  \href {http://adsabs.harvard.edu/abs/1999MNRAS.304..793C} {304, 793}

\bibitem[\protect\citeauthoryear{{Chen} et~al.,}{{Chen} et~al.}{2016}]{Chen16}
{Chen} Y.-T.,  et~al., 2016, \mn@doi [\apjl] {10.3847/2041-8205/827/2/L24},
  \href {http://adsabs.harvard.edu/abs/2016ApJ...827L..24C} {827, L24}

\bibitem[\protect\citeauthoryear{{Giorgini} et~al.,}{{Giorgini}
  et~al.}{1996}]{Giorgini96}
{Giorgini} J.~D.,  et~al., 1996, in AAS/Division for Planetary Sciences Meeting
  Abstracts \#28. p.~1158

\bibitem[\protect\citeauthoryear{{Gladman} et~al.,}{{Gladman}
  et~al.}{2009}]{Gladmanetal09}
{Gladman} B.,  et~al., 2009, \mn@doi [\apjl] {10.1088/0004-637X/697/2/L91},
  \href {http://adsabs.harvard.edu/abs/2009ApJ...697L..91G} {697, L91}

\bibitem[\protect\citeauthoryear{{Heisler} \& {Tremaine}}{{Heisler} \&
  {Tremaine}}{1986}]{Heisler86}
{Heisler} J.,  {Tremaine} S.,  1986, \mn@doi [\icarus]
  {10.1016/0019-1035(86)90060-6}, \href
  {http://adsabs.harvard.edu/abs/1986Icar...65...13H} {65, 13}

\bibitem[\protect\citeauthoryear{{Jewitt}}{{Jewitt}}{2018}]{Jewitt18}
{Jewitt} D.,  2018, \mn@doi [\aj] {10.3847/1538-3881/aaa1a4}, \href
  {http://adsabs.harvard.edu/abs/2018AJ....155...56J} {155, 56}

\bibitem[\protect\citeauthoryear{{Kne{\v{z}}evi\'c} \&
  {Milani}}{{Kne{\v{z}}evi\'c} \& {Milani}}{2012}]{Knezevic12}
{Kne{\v{z}}evi\'c} Z.,  {Milani} A.,  2012, in IAU Joint Discussion. p.~P18

\bibitem[\protect\citeauthoryear{{Kozai}}{{Kozai}}{1962}]{Kozai62}
{Kozai} Y.,  1962, \mn@doi [\aj] {10.1086/108790}, \href
  {http://adsabs.harvard.edu/abs/1962AJ.....67..591K} {67, 591}

\bibitem[\protect\citeauthoryear{{Levison}, {Duncan}, {Brasser}  \&
  {Kaufmann}}{{Levison} et~al.}{2010}]{Levison10}
{Levison} H.~F.,  {Duncan} M.~J.,  {Brasser} R.,   {Kaufmann} D.~E.,  2010,
  \mn@doi [Science] {10.1126/science.1187535}, \href
  {http://adsabs.harvard.edu/abs/2010Sci...329..187L} {329, 187}

\bibitem[\protect\citeauthoryear{{Lidov}}{{Lidov}}{1962}]{Lidov62}
{Lidov} M.~L.,  1962, \mn@doi [\planss] {10.1016/0032-0633(62)90129-0}, \href
  {http://adsabs.harvard.edu/abs/1962P%26SS....9..719L} {9, 719}

\bibitem[\protect\citeauthoryear{{Morais} \& {Namouni}}{{Morais} \&
  {Namouni}}{2013a}]{MoraisNamouni13a}
{Morais} M.~H.~M.,  {Namouni} F.,  2013a, \mn@doi [Celestial Mechanics and
  Dynamical Astronomy] {10.1007/s10569-013-9519-2}, \href
  {http://adsabs.harvard.edu/abs/2013CeMDA.117..405M} {117, 405}

\bibitem[\protect\citeauthoryear{{Morais} \& {Namouni}}{{Morais} \&
  {Namouni}}{2013b}]{MoraisNamouni13b}
{Morais} M.~H.~M.,  {Namouni} F.,  2013b, \mn@doi [\mnras]
  {10.1093/mnrasl/slt106}, \href
  {http://adsabs.harvard.edu/abs/2013MNRAS.436L..30M} {436, L30}

\bibitem[\protect\citeauthoryear{{Morais} \& {Namouni}}{{Morais} \&
  {Namouni}}{2016}]{MoraisNamouni16}
{Morais} M.~H.~M.,  {Namouni} F.,  2016, \mn@doi [Celestial Mechanics and
  Dynamical Astronomy] {10.1007/s10569-016-9674-3}, \href
  {http://adsabs.harvard.edu/abs/2016CeMDA.125...91M} {125, 91}

\bibitem[\protect\citeauthoryear{{Morais} \& {Namouni}}{{Morais} \&
  {Namouni}}{2017}]{MoraisNamouni17b}
{Morais} M.~H.~M.,  {Namouni} F.,  2017, \mnras, \href
  {http://doi.org/10.1093/mnrasl/slx125} {472, L1}

\bibitem[\protect\citeauthoryear{{Namouni} \& {Morais}}{{Namouni} \&
  {Morais}}{2015}]{NamouniMorais15}
{Namouni} F.,  {Morais} M.~H.~M.,  2015, \mn@doi [\mnras]
  {10.1093/mnras/stu2199}, \href
  {http://adsabs.harvard.edu/abs/2015MNRAS.446.1998N} {446, 1998}

\bibitem[\protect\citeauthoryear{{Namouni} \& {Morais}}{{Namouni} \&
  {Morais}}{2017a}]{NamouniMorais17}
{Namouni} F.,  {Morais} M.~H.~M.,  2017a, \mn@doi [\mnras]
  {10.1093/mnras/stx290}, 467, 2673

\bibitem[\protect\citeauthoryear{{Namouni} \& {Morais}}{{Namouni} \&
  {Morais}}{2017b}]{NamouniMorais17b}
{Namouni} F.,  {Morais} M.~H.~M.,  2017b, \mn@doi [\mnras]
  {10.1093/mnras/stx1714}, \href
  {http://adsabs.harvard.edu/abs/2017MNRAS.471.2097N} {471, 2097}

\bibitem[\protect\citeauthoryear{{Namouni} \& {Morais}}{{Namouni} \&
  {Morais}}{2017c}]{NamouniMorais17c}
{Namouni} F.,  {Morais} M.~H.~M.,  2017c, \mn@doi [J. Comp. App. Math.]
  {10.1007/s40314-017-0489-y}, \href
  {http://adsabs.harvard.edu/abs/2017arXiv170400550N} {published online.
  arXiv:1704.00550, http://dx.doi.org/10.1007/s40314}

\bibitem[\protect\citeauthoryear{{Nesvorn{\'y}}, {Vokrouhlick{\'y}}  \&
  {Morbidelli}}{{Nesvorn{\'y}} et~al.}{2013}]{Nesvorny13}
{Nesvorn{\'y}} D.,  {Vokrouhlick{\'y}} D.,   {Morbidelli} A.,  2013, \mn@doi
  [\apj] {10.1088/0004-637X/768/1/45}, \href
  {http://adsabs.harvard.edu/abs/2013ApJ...768...45N} {768, 45}

\bibitem[\protect\citeauthoryear{{Pfalzner} et~al.,}{{Pfalzner}
  et~al.}{2015}]{Pfalzner15}
{Pfalzner} S.,  et~al., 2015, \mn@doi [\physscr]
  {10.1088/0031-8949/90/6/068001}, \href
  {http://adsabs.harvard.edu/abs/2015PhyS...90f8001P} {90, 068001}

\bibitem[\protect\citeauthoryear{{Robutel} \& {Gabern}}{{Robutel} \&
  {Gabern}}{2006}]{Robutel06}
{Robutel} P.,  {Gabern} F.,  2006, \mn@doi [\mnras]
  {10.1111/j.1365-2966.2006.11008.x}, \href
  {http://adsabs.harvard.edu/abs/2006MNRAS.372.1463R} {372, 1463}

\bibitem[\protect\citeauthoryear{{Thomopoulos}}{{Thomopoulos}}{2013}]{cholesky}
{Thomopoulos} N.~T.,  2013, {Essentials of Monte Carlo Simulation}.
{Springer-Verlag New York}

\bibitem[\protect\citeauthoryear{{Tiscareno} \& {Malhotra}}{{Tiscareno} \&
  {Malhotra}}{2003}]{TiscarenoMalhotra03}
{Tiscareno} M.~S.,  {Malhotra} R.,  2003, \mn@doi [\aj] {10.1086/379554}, \href
  {http://adsabs.harvard.edu/abs/2003AJ....126.3122T} {126, 3122}

\bibitem[\protect\citeauthoryear{{Tsiganis}, {Varvoglis}  \&
  {Dvorak}}{{Tsiganis} et~al.}{2005}]{Tsiganis05}
{Tsiganis} K.,  {Varvoglis} H.,   {Dvorak} R.,  2005, \mn@doi [Celestial
  Mechanics and Dynamical Astronomy] {10.1007/s10569-004-3975-7}, \href
  {http://adsabs.harvard.edu/abs/2005CeMDA..92...71T} {92, 71}

\bibitem[\protect\citeauthoryear{{Volk} \& {Malhotra}}{{Volk} \&
  {Malhotra}}{2013}]{VolkMalhotra13}
{Volk} K.,  {Malhotra} R.,  2013, \mn@doi [\icarus]
  {10.1016/j.icarus.2013.02.016}, \href
  {http://adsabs.harvard.edu/abs/2013Icar..224...66V} {224, 66}

\bibitem[\protect\citeauthoryear{{Widmark} \& {Monari}}{{Widmark} \&
  {Monari}}{2017}]{Widmark17}
{Widmark} A.,  {Monari} G.,  2017, preprint, \href
  {http://adsabs.harvard.edu/abs/2017arXiv171107504W} {} (\mn@eprint {arXiv}
  {1711.07504})

\bibitem[\protect\citeauthoryear{{Wiegert} \& {Tremaine}}{{Wiegert} \&
  {Tremaine}}{1999}]{Wiegert99}
{Wiegert} P.,  {Tremaine} S.,  1999, \mn@doi [\icarus]
  {10.1006/icar.1998.6040}, \href
  {http://adsabs.harvard.edu/abs/1999Icar..137...84W} {137, 84}

\bibitem[\protect\citeauthoryear{{Wiegert}, {Connors}  \& {Veillet}}{{Wiegert}
  et~al.}{2017}]{Wiegert17}
{Wiegert} P.,  {Connors} M.,   {Veillet} C.,  2017, Nature, 543, 687

\makeatother
\end{thebibliography}

\end{document}